\documentclass[doublecol]{epl2}
\usepackage{amsmath, bm, amssymb}
\definecolor{linkcolor}{rgb}{0,0,0.6} 
\usepackage[pdftex,colorlinks=true,
	pdfstartview = FitV,
	linkcolor    = linkcolor,
	citecolor    = linkcolor,
	urlcolor     = linkcolor,
	hyperindex   = true,
	hyperfigures = false]{hyperref}

\title{Active engines: Thermodynamics moves forward}

\author{\'Etienne Fodor\inst{1} \and Michael E. Cates\inst{2}}
\shortauthor{\'Etienne Fodor and Michael E. Cates}

\institute{                    
  \inst{1} Department of Physics and Materials Science, University of Luxembourg, L-1511 Luxembourg
	\\
	\inst{2} DAMTP, Centre for Mathematical Sciences, University of Cambridge, Wilberforce Road, Cambridge CB3 0WA, UK
}


\abstract{
	The study of thermal heat engines was pivotal to establishing the principles of equilibrium thermodynamics, with implications far wider than only engine optimization. For nonequilibrium systems, which by definition dissipate energy even at rest, how to best convert such dissipation into useful work is still largely an outstanding question, with similar potential to illuminate general physical principles. We review recent theoretical progress in studying the performances of engines operating with active matter, where particles are driven by individual self-propulsion. We distinguish two main classes, either autonomous engines exploiting a particle current, or cyclic engines applying periodic transformation to the system, and present the strategies put forward so far for optimization. We delineate the limitations of previous studies, and propose some futures perspectives, with a view to building a consistent thermodynamic framework far from equilibrium.
}

\begin{document}

\maketitle

% ===============================================================================

The foundation of equilibrium thermodynamics was largely motivated by the study of thermal engines. During the Industrial Revolution of the 19$^{\rm th}$ century, an important challenge was to establish a theoretical framework to rationalize how to best convert different forms of energy, such as work and heat. The main contributions of Carnot and Stirling was to provide simple protocols, based on a minimal description of engines, for which the performances can be evaluated exactly in terms of a few parameters~\cite{Carnot}. These protocols served as a cornerstone to establish the principles of equilibrium thermodynamics, thus eclipsing their original motivation of optimizing realistic engines.

Thermal engines, which operate by manipulating a passive system, can only extract work by changing its temperature. Interestingly, many nonequilibrium systems dissipate heat despite the absence of any temperature variation. This removes a major constraint, and opens the door to designing innovative protocols with no passive counterpart. Following this route, a main difficulty lies in rationalizing properly the performances of engines which fall beyond the scope of standard thermodynamics. The study of such innovative engines then requires a search for a systematic recipe to evaluate properly, and to exploit efficiently, the energy fluxes of nonequilibrium systems. As holds for thermal engines, this target typically amounts to producing maximum work while dissipating minimum heat. The important outstanding challenge is to delineate the generic principles for the design of nonequilibrium engines, encompassing within a unified framework a large class of systems beyond equilibrium. Whether or not such engines are practically useful, one can hope to gain broader insights by studying them.

In recent decades, active matter has emerged as the class of nonequilibrium systems whose microscopic constituents extract energy from their environment to produce a systematic, autonomous motion~\cite{Marchetti2013, Bechinger2016, Marchetti2018}. Examples of active matter can be found either in (i)~biological systems, such as swarms of bacteria~\cite{Elgeti2015} and cell tissues~\cite{Ladoux2017}, (ii)~social systems, such as groups of animals~\cite{Cavagna2014} and human crowds~\cite{Bartolo2019}, or (iii)~synthetic systems, such as self-catalytic colloids in a fuel bath~\cite{Palacci2013} and vibrated polar particles~\cite{Dauchot2010}. The combination of individual self-propulsion and interactions between individual particles leads to nonequilibrium collective behaviors. Examples include a collective directed motion with long-ranged polar order~\cite{Chate2020}, and a phase separation for purely repulsive constituents~\cite{Cates2015}.

To rationalize these collective behaviors in terms of simple ingredients, minimal models have been proposed. In most cases, the details of microscopic interactions allow one to anticipate which type of order emerges at large scale, thus defining different classes of active systems~\cite{Marchetti2013}. Based on these minimal models, numerous studies have striven to characterize the properties of active matter within a unified framework, by relying mostly on the partial applicability of thermodynamic concepts beyond equilibrium. This includes scrutiny of observables directly inspired by equilibrium studies, such as pressure~\cite{Marchetti2014, Brady2014, Solon2015}, surface tension~\cite{Paliwal2017, Zakine2020}, and chemical potential~\cite{Paliwal2018, Guioth2019}.

\begin{figure}
	\centering
	\includegraphics[width=.75\linewidth]{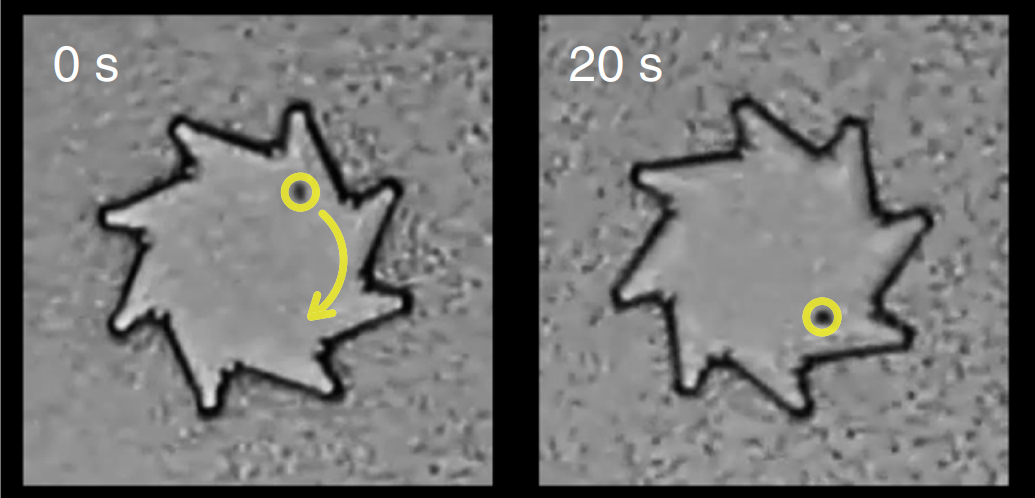}
	\vskip.2cm
	\includegraphics[width=.6\linewidth]{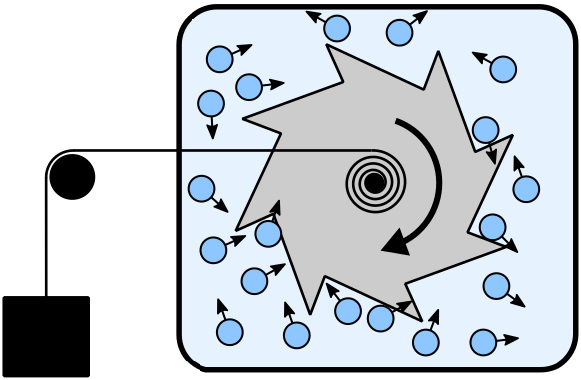}
	\caption{(Color online)~Illustration of autonomous active engines.
		(Top)~Snapshots of an experimental active ratchet, made of a wheel with asymmetric teeth immersed in a bath of bacteria. The breakdown of both time-reversal symmetry, induced by the nonequilibrium dynamics of bacteria, and spatial symmetry, induced by the wheel shape, leads to a spontaneous current, given here by rotation of the wheel. Taken from~\cite{Leonardo2010}.
		(Bottom)~The engine extracts work by applying an external load opposed to the spontaneous current. Here, the load consists in connecting an external mass to the wheel, which enforces a counter-torque opposed to the spontaneous rotation. Adapted from~\cite{Pietzonka2019}.
	}
	\label{fig1}
\end{figure}

Although incomplete, this thermodynamic framework sheds light on many anomalous properties of active matter, allowing one to better delineate its deviation from equilibrium. For instance, introducing an external wall to probe the internal pressure perturbs the system so strongly that the measurement can not be disentangled from the details of the wall: At variance with equilibrium, the pressure does not follow any equation of state in most dry active systems~\cite{Solon2015}. Moreover, embedding asymmetric obstacles in a bath of active particles leads to spontaneous particle currents: See fig.~\ref{fig1}. These ratchets exploit the simultaneous breakdown of spatial symmetry, encoded in the obstacle shape, and of time-reversal symmetry, induced by any nonequilibrium dynamics~\cite{Prost1997}. A large variety of ratchets have been proposed and studied, both experimentally~\cite{Galajda2007, Leonardo2010, Aranson2010} and theoretically~\cite{Tailleur2009, Cates2016, Reichhardt2017, Pietzonka2019}.

Many works have focused on how to characterize the irreversibility of the dynamics even in the absence of obstacles~\cite{Nardini2016, Mandal2017, Nardini2017, Seifert2018, Bo2019, Cates2021}. To this aim, one can rely on the framework of stochastic thermodynamics, which allows one to evaluate irreversibility in terms of some specific observables. This framework was first established in passive systems to extend standard notions of thermodynamics, such as work and heat, to small scales~\cite{Sekimoto1998, Seifert2012}. It offers systematic methods to identify the energy transfers between a system and its environment using a {\it bottom-up} approach, based on analyzing directly the dynamics of the microscopic constituents. Stochastic thermodynamics thereby creates a useful framework to build thermodynamic principles out-of-equilibrium, since it does not require us to define {\it a priori} any macroscopic state function. Various recent studies have shown how to extend such a framework from passive to active systems, thus providing definitions of work and heat in active matter~\cite{Speck2016, Seifert2018, Suri2019, Bo2019, Suri2020, Cates2021}.

Recent developments in active matter and in stochastic thermodynamics have thus led to many important results on the thermodynamic properties of active systems. By now, our understanding of active matter is sufficiently advanced and well-established to rationalize (i)~how to detect unambiguously the nonequilibrium properties of active systems, and (ii)~how to disentangle properly the energy transfers induced either by the microscopic activity or by any external protocol. This knowledge fosters unprecedented hopes for building a systematic, versatile framework of design principles for the reliable control of active systems. Towards such a framework, a first and important step consists in studying a specific class of protocols which extract energy from active matter. In doing so, one motivation is to go beyond simply characterizing nonequilibrium properties, to exploit the deviation from equilibrium to design active engines. A second motivation is the hope that, as happened for equilibrium thermodynamics, study of  specific engines can illuminate general principles.

\begin{figure}
	\centering
	\includegraphics[width=.45\linewidth]{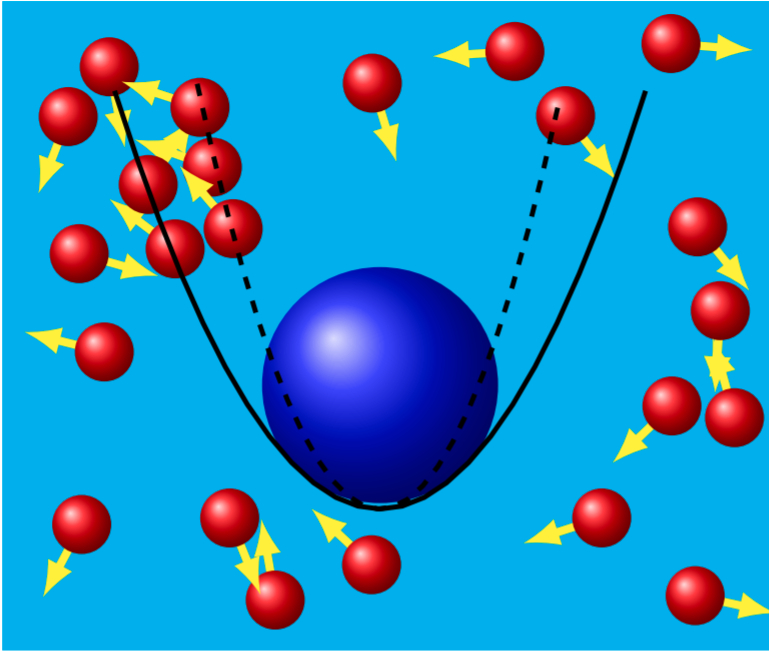}
	\hskip1cm
	\raisebox{-1cm}{\rotatebox{90}{\includegraphics[width=.61\linewidth]{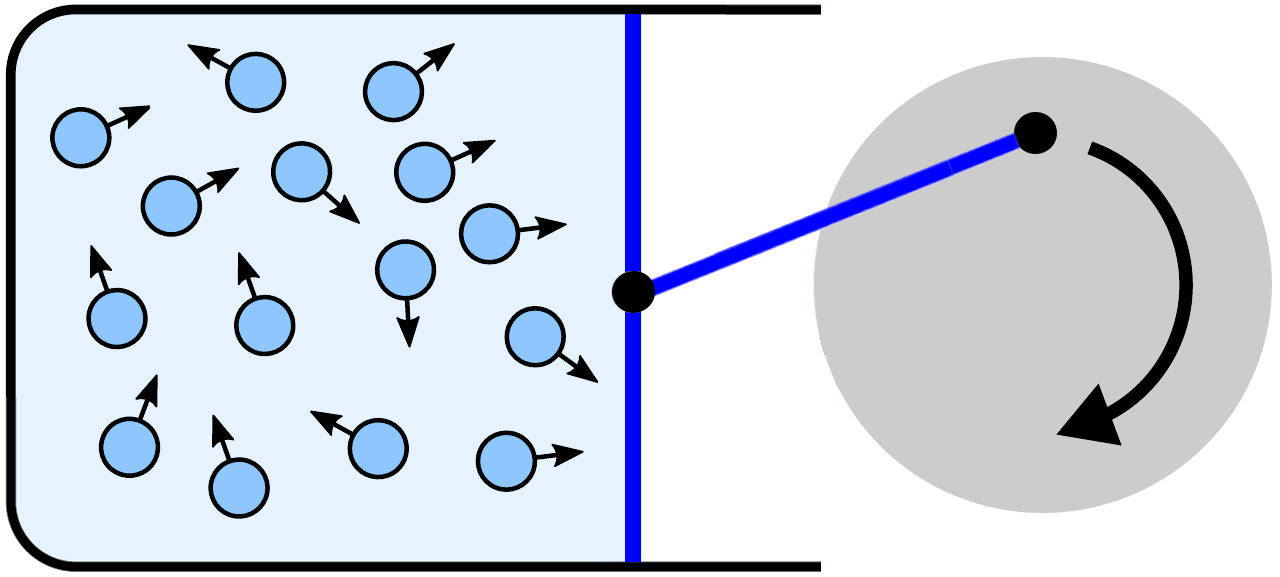}}}
	\caption{(Color online)~Illustration of cyclic active engines.
		(Left)~Colloidal engine made of a passive tracer, in blue, immersed in a bath of active particles, in red, and confined in a harmonic trap whose stiffness can be varied externally. A typical cycle consists in synchronizing the trap stiffness with some parameters of the bath. Taken from~\cite{Holubec2020}.
		(Right)~Macroscopic engine made of active particles confined in a container. The rightmost boundary is connected to an external wheel, hence periodically compressing and expanding the assembly of active particles. A typical cycle consists in synchronizing the volume of the container with either some properties of the boundaries, such as their stiffness, or some properties of the confined particles, such as their temperature or activity.
	}
	\label{fig2}
\end{figure}

In what follows, we present a short review of various active engines proposed in the literature. The challenge is to search for appropriate protocols which convert the random motion of self-propelled particles into extractable work, and to establish a systematic road-map for the optimization of engine performances. Some protocols are inspired by experimental realizations, others rely on innovative settings which might motivate future experiments. We consider minimal models of self-propelled particles, which contain the essential ingredients at play in active matter, and we discuss how to exploit properly their nonequilibrium properties. We distinguish two main classes of active engines: (i)~autonomous engines, which exploit the ratchet currents, see fig.~\ref{fig1}, and (ii)~cyclic engines, which operate by applying periodic transformation to active and/or passive particles, see fig.~\ref{fig2}. Within each class, we present some simple examples, which could inspire more complex ones, and we characterize their performances with universal observables, namely efficiency and power.

% -------------------------------------------------------------------------------

{\bf Particle-based description of active engines.}~--~We consider a generic setting where the engine can be described in terms of (i)~the coordinates of some passive elements $\{{\bf x}_a\}$, for instance the positions of obstacles, and (ii)~a set of active coordinates $\{{\bf r}_i, \theta_i\}$, typically the position and orientation of active particles in two spatial dimensions. Provided that inertial effects are negligible, their dynamics is given by coupled, overdamped Langevin equations. Each of these encodes a balance between the forces stemming from a thermostat, namely viscous drag and thermal noise, that deriving from external potential, and some non-conservative forces. Introducing the potential $U(\{{\bf x}_a,{\bf r}_i,\theta_i\})$, the dynamics of passive elements reads
\begin{equation}\label{eq:dyn_pas}
	\dot{\bf x}_a = \mu_p \bigg( {\bf f}_a - \frac{\partial U}{\partial{\bf x}_a}\bigg) + \sqrt{2D_p}{\bm\zeta}_a ,
\end{equation}
and that of active particles is (with ${\bf e}_i = (\cos\theta_i,\sin \theta_i)$)
\begin{equation}
	\begin{aligned}\label{eq:dyn_act}
		\dot{\bf r}_i &= v {\bf e}_i + \mu \bigg( {\bf f}_i - \frac{\partial U}{\partial{\bf r}_i}\bigg) + \sqrt{2D}{\bm\xi}_i ,
		\\
		\dot\theta_i &= \mu_r \bigg(\omega_i - \frac{\partial U}{\partial\theta_i}\bigg) + \sqrt{2D_r} \eta_i .
	\end{aligned}
\end{equation}
The terms $\{{\bf f}_a, {\bf f}_i, \omega_i\}$ represent non-conservative forces and torques which can in principle be applied on the various elements of the system by an external agent or operator. The active force $v {\bf e}_i$ has a constant norm $v$, and its orientation is given by the unit orientation vector ${\bf e}_i$. This force represents the conversion of some energy fuel present in the environment, for instance ATP in living systems~\cite{Ladoux2017}, into an autonomous self-propulsion. The terms $\{{\bm\zeta}_a, {\bm\xi}_i, \eta_i\}$ are uncorrelated Gaussian white noises, with zero mean and unit variance. The diffusion coefficients $\{D_p, D, D_r\}$ and the mobilities $\{\mu_p, \mu, \mu_r\}$ are generally constrained so that their respective ratio is set by the temperature $T$ of the surrounding thermostat: $T = D_p/\mu_p = D/\mu = D_r/\mu_r$. Note that these equations do not conserve momentum and therefore represent a ``dry'' system, such as a 2D layer of active particles on a solid support with which momentum is exchanged.

Importantly, we take the potential $U$ to be under external control, as captured by a set of parameters $\{\alpha_k\}$ which determine the shape of $U$. By changing $\{\alpha_k\}$ as a function of time, the operator can vary the shape of an external potential applied on each particle (or a subset of them), and also in principle control how particles interact between themselves. For autonomous engines, the operator typically applies some constant forces and/or torques $\{{\bf f}_a, {\bf f}_i, \omega_i\}$ which are opposed to the spontaneous current of the corresponding passive and/or active coordinates, without changing $U$, see fig.~\ref{fig1}. In contrast, for cyclic engines, the parameters $\{\alpha_k\}$ are varied periodically in time without applying any external force or torque, see fig.~\ref{fig2}. While both features could be combined in a single device, this division into two basic types is conceptually helpful.

% -------------------------------------------------------------------------------

{\bf Extracted work and dissipated heat.}~--~Following standard definitions of stochastic thermodynamics~\cite{Sekimoto1998, Seifert2012}, the work $\cal W$ produced during a protocol time $\tau$ is
\begin{equation}\label{eq:work}
	{\cal W} = \int_0^\tau dt \bigg[ \sum_k \dot \alpha_k \frac{\partial U}{\partial\alpha_k} + \sum_a {\bf f}_a\cdot\dot{\bf x}_a + \sum_i \big( {\bf f}_i\cdot\dot{\bf r}_i + \omega_i \dot\theta_i \big) \bigg] .
\end{equation}
The sign convention is that energy is extracted from the system whenever ${\cal W}<0$. The work $\cal W$ is a stochastic process which fluctuates from one realization of the protocol to another. The challenge in designing an engine is to ensure that the work remains negative {\it on average}. Importantly, the definition in~\eqref{eq:work} only involves terms describing how the operator perturbs the dynamics. It is agnostic to the presence of any other non-conservative force, and thus~\eqref{eq:work} is the same for a passive system.

To evaluate the amount of heat $\cal Q$ dissipated by the degrees of freedom $\{{\bf x}_a,{\bf r}_i,\theta_i\}$, it suffices to quantify the power of the forces and torques exerted by the thermostat on each one of them~\cite{Sekimoto1998, Seifert2012}. Indeed, by definition, the thermostat can only exchange energy with the system in the form of heat. Therefore, given the viscous drag terms and the thermal noises in~(\ref{eq:dyn_pas}-\ref{eq:dyn_act}), the heat follows as
\begin{equation}
	\begin{aligned}\label{eq:heat}
		&{\cal Q} = \int_0^\tau dt \bigg\{ \sum_a \frac{\dot{\bf x}_a}{\mu_p} \cdot \big( \dot{\bf x}_a - \sqrt{2D_p}{\bm\zeta}_a \big)
		\\
		&\, + \sum_i \bigg[ \frac{\dot{\bf r}_i}{\mu} \cdot \big( \dot{\bf r}_i - \sqrt{2D}{\bm\xi}_i \big) + \frac{\dot\theta_i}{\mu_r} \big( \dot\theta_i - \sqrt{2D_r}\eta_i \big) \bigg] \bigg\} .
	\end{aligned}
\end{equation}
The heat $\cal Q$ is a stochastic process, like the work $\cal W$, and is defined such that energy is transferred {\it from} the system {\it to} the thermostat whenever ${\cal Q}>0$. The definition in~\eqref{eq:heat} again holds even for a passive system ($v=0$). Substituting the dynamics~(\ref{eq:dyn_pas}-\ref{eq:dyn_act}) into eq.~\eqref{eq:heat}, we deduce
\begin{equation}\label{eq:heat_bis}
	\begin{aligned}
	&{\cal Q} = \int_0^\tau dt \bigg\{\sum_a \bigg( {\bf f}_a - \frac{\partial U}{\partial{\bf x}_a}\bigg)\cdot\dot{\bf x}_a
	\\
 	&\,+ \sum_i \bigg[ \bigg( \frac{v{\bf e}_i}{\mu} + {\bf f}_i - \frac{\partial U}{\partial{\bf r}_i}\bigg)\cdot\dot{\bf r}_i + \bigg(\omega_i - \frac{\partial U}{\partial\theta_i}\bigg)\dot\theta_i \bigg]\bigg\} .
	\end{aligned}
\end{equation}
For constant $\{\alpha_k\}$ and in the absence of $\{{\bf f}_a, {\bf f}_i, \omega_i\}$, the external operator does not perturb the dynamics. The average heat is then extensive in time: $\langle{\cal Q}\rangle = (v\tau/\mu) \sum_i \langle {\bf e}_i\cdot\dot{\bf r}_i \rangle$, which illustrates that there is always a constant rate of energy dissipated by the active forces, independently of any external protocol. Finally, combining the expression of work and heat in~\eqref{eq:work} and~\eqref{eq:heat_bis}, we get
\begin{equation}\label{eq:1st_law}
	U(t=\tau) - U(t=0) = {\cal W} - {\cal Q} + \int_0^\tau dt \sum_i \frac{v {\bf e}_i}{\mu} \cdot \dot{\bf r}_i .
\end{equation}
For passive systems ($v=0$), this is the first law of thermodynamics (FLT), as expected. For an active system, the energy budget written in~\eqref{eq:1st_law} no longer involves solely the potential $U$, the work $\cal W$, and the heat $\cal Q$. Crucially, it now also accounts for the energy spent sustaining the self-propulsion of active particles.

While the definition of work, given in~\eqref{eq:work}, is standard in most of the literature, how heat should be properly defined remains widely debated. One reason for the controversy is that the thermal noises in the dynamics~(\ref{eq:dyn_pas}-\ref{eq:dyn_act}), are sometimes omitted in active matter studies. In this context, some works have defined heat by replacing the thermal noises appearing in~\eqref{eq:heat} with the corresponding active forces for each degree of freedom~\cite{Zakine2017, Park2020, Roldan2020}. This amounts to enforcing that the standard FLT, {\it i.e.} eq.~\eqref{eq:1st_law} with $v=0$, should hold in the same form for active and passive systems. Even when the dynamics features thermal noises, other works have defined heat by substituting the sum of thermal noises and active forces, instead of thermal noises only, in~\eqref{eq:heat} \cite{Holubec2020, Saha2020}. Again, this yields the standard FLT.

Importantly, the main drawback of these definitions of heat, leading to the standard FLT, is that they do not capture the energy dissipated by active forces. Indeed, when the operator does not perturb the dynamics, these definitions lead to a vanishing average heat, which suggests misleadingly that the self-propulsion can be sustained at zero cost. To avoid such an inconsistency, one should formally distinguish the thermal noises from the active forces, as is done in~\eqref{eq:heat}. This applies even when (as is often true in active systems) thermal effects are negligible. Indeed, the self-propulsion stems from the underlying consumption of fuel energy which, although affected by thermal fluctuations, is mainly not of thermal origin.

Yet another way of defining heat imagines the active velocity $v{\bf e}_i$ to stem from a local advective fluid flow, with the displacement of a given active particle evaluated with respect to this flow~\cite{Speck2018, Bo2019}. In this approach, the velocity $\dot{\bf r}_i$ appearing in~\eqref{eq:heat} is replaced systematically with $\dot{\bf r}_i - v{\bf e}_i$, which leads to substituting $\partial U/\partial{\bf r}_i$ instead of $\dot{\bf r}_i/\mu$ in \eqref{eq:1st_law}. With this definition, the average heat in the absence of external perturbation reads $v\tau \sum_i \langle {\bf e}_i\cdot(\partial U / \partial {\bf r}_i) \rangle$, which vanishes when there is no potential $U$. Overall, among the various definitions of heat presented above, the only one which retains a finite cost of energy even in the absence of potential $U$ is the one we have adopted in~\eqref{eq:heat}.

% -------------------------------------------------------------------------------

{\bf Efficiency and power.}~--~The efficiency of the engine $\cal E$ compares the output work with the energetic cost of operating the engine:
\begin{equation}\label{eq:eff}
	{\cal E} = \frac{\langle{\cal W}\rangle}{\langle{\cal W}\rangle - \langle{\cal Q}\rangle} .
\end{equation}
The average heat being always positive, due to the irreversibility of the dynamics, the efficiency is always smaller than unity, as it should be. The definition in~\eqref{eq:eff} is standard for protocols at constant temperature~\cite{Prost1997, Pietzonka2019, Ekeh2020, Szamel2020, Stark2020}. For cycles which vary the temperature, other definitions inspired by that of thermal engines are sometimes preferred, with a view to comparing the performances of thermal cycles with and without activity~\cite{Saha2019, Holubec2020, Holubec2020a, Park2020, Saha2020}. In this context, the definition of heat which is often adopted discards the cost of self-propulsion on purpose, so as to account only for that of changing the bath temperature.

Under steady cycling of a periodic protocol with period $\tau$, the $U$ terms in \eqref{eq:1st_law} cancel on average, so that by substituting~\eqref{eq:1st_law} into~\eqref{eq:eff}, we get
\begin{equation}\label{eq:eff_bis}
	{\cal E} = |\langle{\cal W}\rangle| \,\bigg( \int_0^\tau dt \sum_i \Big\langle \frac{v {\bf e}_i}{\mu} \cdot \dot{\bf r}_i \Big\rangle \bigg)^{-1} .
\end{equation}
This makes it clear that, with our definition, the effective cost of operating an engines only comes from the active forces. Note that this approach deliberately neglects the energetic cost of converting fuel (or another ambient energy supply) into self-propulsion. Indeed, we only resolve the energy budget from the propulsion scale upwards, but not at smaller scales. Accordingly the dynamics~(\ref{eq:dyn_pas}-\ref{eq:dyn_act}) makes no attempt to describe how fuel-consuming processes bring about the active contribution $v{\bf e}_i$ to $\dot{\bf r}_i$.

Whenever the dynamics of active particles is mostly dominated by free motion, we have $\langle {\bf e}_i\cdot\dot{\bf r}_i \rangle \approx v$ in \eqref{eq:eff_bis}. This holds typically when (i)~interactions with passive and/or active elements are negligible, namely for dilute systems, and (ii)~any external force and/or external potential only affects a sub-extensive number of active particles. This encompasses, for instance, autonomous engines whose load is applied only on passive elements, and cyclic engines whose varying potential is located at boundaries of the system. In such cases, it follows from~\eqref{eq:eff_bis} that the efficiency is directly proportional to the extracted power:
\begin{equation}\label{eq:eff_pow}
	{\cal E} \approx \frac{\mu}{Nv^2} \frac{|\langle{\cal W}\rangle|}{\tau} \equiv \frac{\mu\langle{\cal P}\rangle}{Nv^2} ,
\end{equation}
where $N$ denotes here the number of active particles. This important result implies that these cases avoid any trade-off between efficiency and power. In other words, one can address simultaneously the optimization of (i)~how to convert efficiently the random self-propulsion into extractable work, and (ii)~how to extract maximum power from the given active system. This starkly contrasts with the case of thermal cyclic engines, where the efficiency is always maximized by long, quasi-reversible cycle times, for which the power generally vanishes~\cite{Seifert2008, Esposito2009}.

Moreover, for active engines which satisfy the efficiency-power relation~\eqref{eq:eff_pow}, the fluctuations of the power obey the following bound~\cite{Ekeh2020}:
\begin{equation}\label{eq:bound}
	\frac{1}{\langle{\cal P}^2\rangle - \langle{\cal P}\rangle^2} \bigg[\bigg( 1 + \tau\frac{d}{d\tau}\bigg) \langle{\cal P}\rangle\bigg]^2 \leq \frac{\mu\tau}{2D} \bigg( \frac{N v^2}{\mu} - \langle{\cal P}\rangle \bigg) .
\end{equation}
The derivation of this bound builds on recent studies of thermodynamic uncertainty relations~\cite{Gingrich2016, Koyuk2019}. Equation~\eqref{eq:bound} suggests that, when the average power approaches $Nv^2/\mu$, or equivalently when the efficiency gets close to unity, the fluctuations of the power become very large. Indeed, the bound requires the variance $\langle{\cal P}^2\rangle - \langle{\cal P}\rangle^2$ to be at least of order $\langle{\cal P}\rangle^2$ in this limit. However, the bound \eqref{eq:bound} is established only in regimes where~\eqref{eq:eff_pow} should hold. As described above, this requires the active particles to spend most of their time dissipating energy by moving idly, without contributing much to the extracted work. This corresponds typically to low efficiency, with the power being far smaller than $N v^2/\mu$, weakening the bound in \eqref{eq:bound}. It is an open question to what extent~(\ref{eq:eff_pow}-\ref{eq:bound}) remain valid beyond this ``weak work extraction'' regime.

% -------------------------------------------------------------------------------

{\bf Autonomous engines: Ratchets under load.}~--~An important class of active engines are those made of ratchets~\cite{Prost1997}, where asymmetric potentials lead to a current of active and/or passive particles. These {\it autonomous engines} extract work by applying an external load, typically either a constant force or a constant torque, opposed to the spontaneous current. Examples of ratchets in experimental active systems, which operate in general by rectifying the random active fluctuations (mainly in living organisms so far), include various shapes of obstacles~\cite{Bechinger2016}. Examples include gear-wheels with asymmetric teeth, which spontaneously rotate~\cite{Leonardo2010, Aranson2010}, see fig.~\ref{fig1}; fixed chevrons, which establish density differences between spatial compartments~\cite{Galajda2007}; and micro-patterns on a substrate which create asymmetric potential barriers guiding the migration of cells~\cite{Mahmud2009}. These experiments have inspired many theoretical works on active ratchets~\cite{Reichhardt2017}. Some of them consider either dilute regimes or one-body problems without any particle interaction~\cite{Tailleur2009, Maggi2014, Yariv2014}, yielding explicit analytical predictions~\cite{Martin2020}. Other studies include some form of particle interactions, allowing a more quantitative comparison with experiments~\cite{Reichhardt2008, Angelani2009}.

\begin{figure}
	\centering
	\includegraphics[width=.49\linewidth]{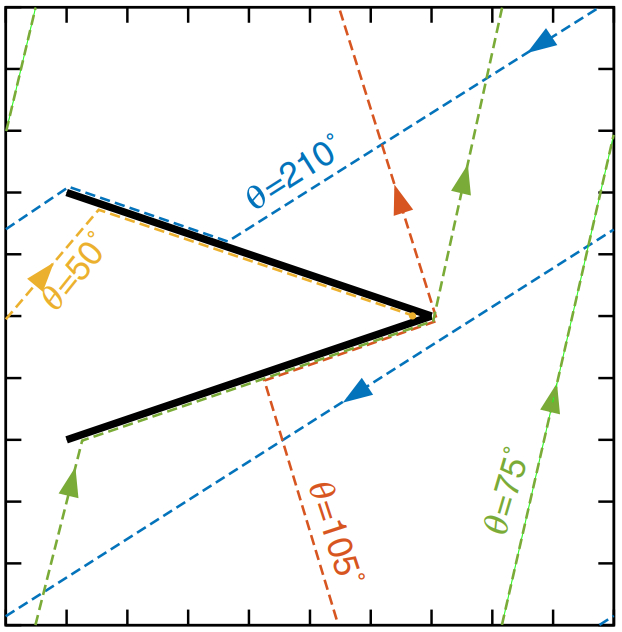}
	\includegraphics[width=.49\linewidth]{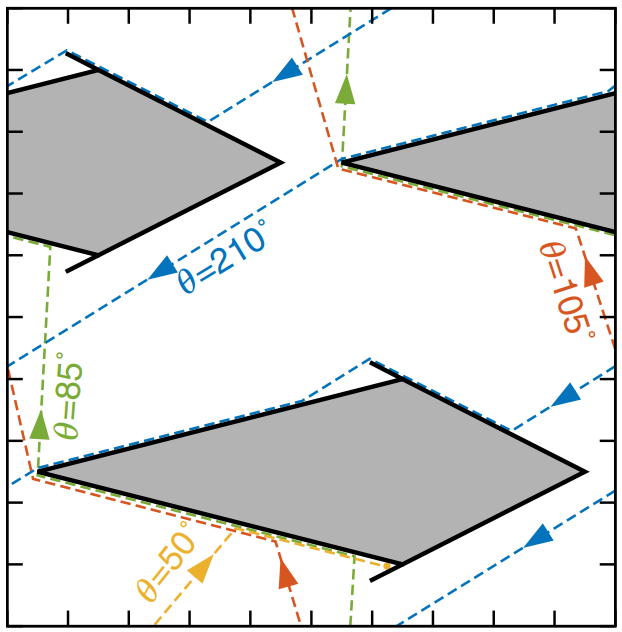}
	\caption{(Color online)~Optimization of an autonomous active engine. Asymmetric obstacles are in contact with a bath of active particles, not shown here, which leads to a spontaneous motion of the obstacles to the right under periodic boundary conditions. The trajectories of particles at high persistence are shown as dashed straight lines for several impact angles $\theta$.
		(Left)~For chevron-shaped obstacles, all trajectories below a certain angle get trapped, see $\theta=50^\circ$, and thus contribute to positive current. Others hinder systematically the displacement of the obstacles, see $\theta=210^\circ$.
		(Right)~For kite-shaped obstacles, some of the trajectories which are not trapped in the wedges still contribute to positive current by gliding over the tail, see $\theta=105^\circ$. Therefore, the overall current of obstacles, given as the sum of currents for all angles, is higher than that of chevron-shaped obstacles.
		Taken from~\cite{Pietzonka2019}.
	}
	\label{fig3}
\end{figure}

Although such ratchets have been studied extensively, only a few works have considered the effect of applying an external load~\cite{Reichhardt2013}, with a view to extracting work~\cite{Pietzonka2019, Roldan2020}. The performances of these engines are characterized by the loading curves of power and efficiency as functions of the external force $f$. Above a critical loading, the spontaneous current is reversed by the external force, in which case the operator provides energy to the system instead of extracting it. Hence, the engine operates properly only within a finite range of loading below the stalling force: $f\in[0,f_s]$. For dilute systems, such as active particles that encounter an assembly of asymmetric obstacles~\cite{Reichhardt2013, Pietzonka2019}, the stalling force is typically of the order of the individual active force $v/\mu$, or smaller. In this case, the engine typically operates close to the regime of linear response, where the current is linear in the deviation from the stalling force: $\langle\dot x \rangle\sim f_s-f$. In contrast, at higher density clogging effects can potentially lead the system far from this regime~\cite{Reichhardt2018}. Within linear response, the power is then simply a parabolic function of the load: $\langle{\cal P}\rangle = f \langle\dot x\rangle \sim f (f_s-f)$, and it has a well defined maximum at $f=f_s/2$. At sufficiently small density, the efficiency is therefore proportional to power, see~\eqref{eq:eff_pow}, and both are maximal at the same loading point.

In this linear regime, the optimization of the engine performances consists essentially in adjusting the shape of obstacles and their spatial distribution in the system, see fig.~\ref{fig3}, to increase the current as much as possible. Indeed, it is sufficient to optimize the current at zero load to increase systematically the maxima of power and efficiency. When the engine consists of obstacles with enforced synchronized motion (as implemented computationally via periodic boundary conditions, fig.~\ref{fig3}) this amounts to treating a purely geometrical problem in the shared co-moving frame of all obstacles~\cite{Pietzonka2019}. For instance, starting from chevrons made of single wedge, it was found empirically that the current is significantly increased by putting an additional wedge at the tail, see fig.~\ref{fig3}, thus forming kite-shaped obstacles. The angles and sizes of the two wedges, and also the distance between each kite, can then be adjusted to optimize simultaneously the maximum power and efficiency. This illustrates how optimizing autonomous engines can guide the design of innovative ratchets, with a view to inspiring future experiments. In more complex settings, machine learning algorithms might help the optimization of obstacle shape, by exploring efficiently a high-dimensional space of geometric parameters.

% -------------------------------------------------------------------------------

{\bf Cyclic engines: Beyond thermal cycles.}~--~Another class of active engines, which do not rely on exploiting a rectified current of passive and/or active particles, operate by applying periodic transformations to the system. These {\it cyclic engines} are inspired by thermal engines, where one changes the temperature periodically~\cite{Carnot}. Note, however, that active cycles can potentially extract work even at constant temperature. Several years ago, a pioneering experiment~\cite{Sood2016} realized a cyclic active engine. This was motivated by colloidal thermal engines comprising a single passive colloid confined in a harmonic trap by optical tweezers~\cite{Blickle2011, Martinez2016}. The active counterpart of~\cite{Sood2016} puts the colloid in contact with a bath of bacteria, see fig.~\ref{fig2}. Here, changing the temperature not only affects the fluctuations coming from the heat bath, but also that stemming from the bath of active particles, whose propulsion is affected by temperature changes. Interestingly, recent experimental techniques have demonstrated the ability to vary the self-propulsion of living organisms by external illumination~\cite{Leonardo2017, Arlt2018}, which opens the door to cycles (possibly but not necessarily monothermal) where activity is controlled independently of temperature.

The experiments outlined above have motivated a large number of theoretical studies on cyclic active engines. Most consider colloidal engines made of a single active particle~\cite{Zakine2017, Saha2019, Holubec2020, Holubec2020a, Saha2020, Szamel2020}, but a few address the case of many-body systems~\cite{Martin2018, Ekeh2020, Stark2020}. Following the experiment in ref.~\cite{Sood2016}, many of these works studied thermal cycles~\cite{Zakine2017, Martin2018, Holubec2020, Holubec2020a, Saha2020, Park2020}, essentially to determine whether adding activity can lead the system to outperform thermal engines. As discussed following \eqref{eq:1st_law} above, one difficulty in addressing this issue is to extend properly the definition of heat beyond the case of passive systems. Importantly, the results of~\cite{Sood2016} rely on assuming that the FLT has the same form in passive and active systems. Although this approach allows one to estimate heat without measuring explicitly the power of active forces, it also disregards the energy cost of sustaining activity of the system. How to estimate the  dissipation induced by active forces in experiments, which would typically require measuring directly these nonequilibrium fluctuating forces, is still open.

\begin{figure}
	\centering
	\includegraphics[width=\linewidth, trim=1.8cm 22.4cm 11cm 1.9cm, clip=true]{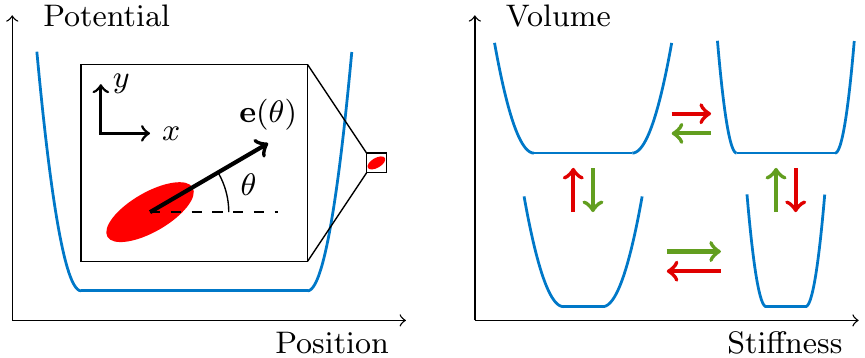}
	\caption{(Color online)~Optimization of a cyclic active engine.
		(Left)~Active elliptical particles, in red, are confined between two parallel soft walls, in blue. The protocol consists in synchronous variation of the distance between walls, which determines the available volume, and their stiffness. Thus, the cycle extracts work by controlling the system only at boundaries, without changing any microscopic property of the confined particles.
		(Right)~Depending on microscopic details of the particles, such as their activity and their anisotropy, the cycle should be operated either clockwise or counter-clockwise. Optimizing the engine performances then requires consideration of the possible cyclic paths in the space of control parameters, and of how the time-dependent execution of such a path affects its power and efficiency.
		Taken from~\cite{Ekeh2020}.
	}
	\label{fig4}
\end{figure}

Interestingly, a number of recent theoretical studies have proposed cycles without any equilibrium counterpart~\cite{Saha2019, Martin2018, Ekeh2020, Szamel2020, Stark2020}, typically operating at constant temperature. These protocols exploit specific nonequilibrium properties of active systems. For instance, given that the pressure exerted by confining walls generally depends on microscopic details of the walls~\cite{Solon2015}, in contrast with equilibrium, one can now vary the pressure at fixed volume and temperature by simply changing some microscopic property of the boundaries, such as their stiffness. Following this line, a protocol has been proposed which, inspired by the Stirling cycle, replaces the changes of temperature with cyclic changes of the stiffness of the walls~\cite{Ekeh2020}. It extracts work by changing volume at fixed stiffness, and changing stiffness at fixed volume, see fig.~\ref{fig4}. Another instance of monothermal cycle in nonequilbrium bath is~\cite{Wulfert2017}.

The performances of these engines are quantified by their power and efficiency as functions of cycle time. For cycle times shorter than a characteristic value, the engine operates too fast to follow properly the instructions of the operator, and does not yield useful work. As the cycle time increases, the power has a non-monotonic behavior; as for thermal cycles, it vanishes at long cycle times and so is maximal in an intermediate regime. In contrast to thermal cycles, the heat expenditure during a cycle contains terms proportional to its duration, due to the dissipation rate induced by active forces. This leads the efficiency to vanish in the quasistatic regime, see~\eqref{eq:eff_bis}. Therefore, since the efficiency is maximal at  a finite cycle time, increasing this maximum efficiency by improved cycle design now requires understanding of non-quasistatic protocols. Developing systematic methods to address the optimization of finite-time cyclic protocols remains an outstanding challenge, which clearly calls for further investigations.

% -------------------------------------------------------------------------------

{\bf Perspective: Exploiting collective effects.}~--~Most of the active engines described above are framed as one-body systems, or (equivalently) systems in which interactions among the components are negligible. It remains largely an open research topic to explore the role of many-body effects in active engines and their consequences for the emerging thermodynamic framework. This motivates close investigation of protocols designed to exploit specific collective effects in active matter. For instance, previous studies have investigated how asymmetric obstacles affect the emergence of flocking in aligning particles~\cite{Reichhardt2012, Valeriani2020}, thus paving the way to designing autonomous engines with such particles as the working substance. As another example, one could study cycles which lead to phase transitions of the confined active particles, such as a phase separation~\cite{Cates2015}. It would be interesting to explore how these transitions affect the engine efficiency and power output.

An alternative way to address how engines might harness emergent order, is to consider hydrodynamic descriptions of activity instead of particle-based dynamics. Previous investigations of ratchet currents in the framework of continuum active fields~\cite{Cates2016, Yeomans2016} open the door to studying autonomous engines at hydrodynamic level. It would be interesting to explore work extraction using cyclic protocols within, for instance, a scalar field theory capturing active phase separation~\cite{Cates2015}, or coupled polar and scalar fields describing flocking~\cite{Chate2020}. A major challenge, in addressing the performances of such engines, is to evaluate the heat dissipated at hydrodynamic level. To this end, one can build on a recent work which proposes a systematic derivation of the energetic costs of active field theories, by embedding these within a framework that encodes the chemical energy flows required to create the activity~\cite{Markovich2020}.

% -------------------------------------------------------------------------------

\acknowledgements{
	The authors acknowledge insightful discussions with Timothy Ekeh, Timur Koyuk, Patrick Pietzonka, and Udo Seifert. \'EF acknowledges support from an ATTRACT Grant of the Luxembourg National Research Fund. MEC is funded by the Royal Society. This work was funded in part by the National Science Foundation, Grant No. NSF PHY-1748958, and the European Research Council under the Horizon 2020 Programme, Grant No. 740269.
}

% ===============================================================================

\bibliographystyle{eplbib}
\bibliography{References}

\begin{thebibliography}{10}
\expandafter\ifx\csname url\endcsname\relax\def\url#1{\texttt{#1}}\fi

\bibitem{Carnot}
\Name{Carnot S.} \Book{R\'eflexions sur la puissance motrice du feu}
  (Bachelier, Paris) 1824.

\bibitem{Marchetti2013}
\Name{Marchetti M.~C. \etal} \REVIEW{Rev. Mod. Phys.}{85}{2013}{1143}.

\bibitem{Bechinger2016}
\Name{Bechinger C. \etal} \REVIEW{Rev. Mod. Phys.}{88}{2016}{045006}.

\bibitem{Marchetti2018}
\Name{Fodor E. \etal} \REVIEW{Physica A}{504}{2018}{106}.

\bibitem{Elgeti2015}
\Name{Elgeti J. \etal} \REVIEW{Rep. Prog. Phys.}{78}{2015}{056601}.

\bibitem{Ladoux2017}
\Name{Saw T.~B. \etal} \REVIEW{Nature}{544}{2017}{212}.

\bibitem{Cavagna2014}
\Name{Cavagna A. \etal} \REVIEW{Annu. Rev. Condens. Matter
  Phys.}{5}{2014}{183}.

\bibitem{Bartolo2019}
\Name{Bain N. \etal} \REVIEW{Science}{363}{2019}{46}.

\bibitem{Palacci2013}
\Name{Palacci J. \etal} \REVIEW{Science}{339}{2013}{936}.

\bibitem{Dauchot2010}
\Name{Deseigne J. \etal} \REVIEW{Phys. Rev. Lett.}{105}{2010}{098001}.

\bibitem{Chate2020}
\Name{Chat\'e H.} \REVIEW{Annu. Rev. Condens. Matter Phys.}{11}{2020}{189}.

\bibitem{Cates2015}
\Name{Cates M.~E. \etal} \REVIEW{Annu. Rev. Condens. Matter
  Phys.}{6}{2015}{219}.

\bibitem{Marchetti2014}
\Name{Yang X. \etal} \REVIEW{Soft Matter}{10}{2014}{6477}.

\bibitem{Brady2014}
\Name{Takatori S.~C. \etal} \REVIEW{Phys. Rev. Lett.}{113}{2014}{028103}.

\bibitem{Solon2015}
\Name{Solon A.~P. \etal} \REVIEW{Nat. Phys.}{11}{2015}{673}.

\bibitem{Paliwal2017}
\Name{Paliwal S. \etal} \REVIEW{J. Chem. Phys.}{147}{2017}{084902}.

\bibitem{Zakine2020}
\Name{Zakine R. \etal} \REVIEW{Phys. Rev. Lett.}{124}{2020}{248003}.

\bibitem{Paliwal2018}
\Name{Paliwal S. \etal} \REVIEW{New J. Phys.}{20}{2018}{015003}.

\bibitem{Guioth2019}
\Name{Guioth J. \etal} \REVIEW{J. Chem. Phys.}{150}{2019}{094108}.

\bibitem{Prost1997}
\Name{J\"ulicher F. \etal} \REVIEW{Rev. Mod. Phys.}{69}{1997}{1269}.

\bibitem{Galajda2007}
\Name{Galajda P. \etal} \REVIEW{J. Bacter.}{189}{2007}{8704}.

\bibitem{Leonardo2010}
\Name{Di~Leonardo R. \etal} \REVIEW{Proc. Natl. Acad. Sci.
  U.S.A.}{107}{2010}{9541}.

\bibitem{Aranson2010}
\Name{Sokolov A. \etal} \REVIEW{Proc. Natl. Acad. Sci. U.S.A.}{107}{2010}{969}.

\bibitem{Tailleur2009}
\Name{Tailleur J. \etal} \REVIEW{{EPL}}{86}{2009}{60002}.

\bibitem{Cates2016}
\Name{Stenhammar J. \etal} \REVIEW{Sci. Adv.}{2}{2016}{e1501850}.

\bibitem{Reichhardt2017}
\Name{Reichhardt C.~O. \etal} \REVIEW{Annu. Rev. Condens. Matter
  Phys.}{8}{2017}{51}.

\bibitem{Pietzonka2019}
\Name{Pietzonka P. \etal} \REVIEW{Phys. Rev. X}{9}{2019}{041032}.

\bibitem{Nardini2016}
\Name{Fodor E. \etal} \REVIEW{Phys. Rev. Lett.}{117}{2016}{038103}.

\bibitem{Mandal2017}
\Name{Mandal D. \etal} \REVIEW{Phys. Rev. Lett.}{119}{2017}{258001}.

\bibitem{Nardini2017}
\Name{Nardini C. \etal} \REVIEW{Phys. Rev. X}{7}{2017}{021007}.

\bibitem{Seifert2018}
\Name{Pietzonka P. \etal} \REVIEW{J. Phys. A: Math. Theor.}{51}{2018}{01LT01}.

\bibitem{Bo2019}
\Name{Dabelow L. \etal} \REVIEW{Phys. Rev. X}{9}{2019}{021009}.

\bibitem{Cates2021}
\Name{Fodor E. \etal} \REVIEW{arXiv e-print}{}{2021}{2104.06634}.

\bibitem{Sekimoto1998}
\Name{Sekimoto K.} \REVIEW{Prog. Theor. Phys. Supp.}{130}{1998}{17}.

\bibitem{Seifert2012}
\Name{Seifert U.} \REVIEW{Rep. Prog. Phys.}{75}{2012}{126001}.

\bibitem{Speck2016}
\Name{Speck T.} \REVIEW{EPL}{114}{2016}{30006}.

\bibitem{Suri2019}
\Name{Tociu L. \etal} \REVIEW{Phys. Rev. X}{9}{2019}{041026}.

\bibitem{Suri2020}
\Name{Fodor E. \etal} \REVIEW{New J. Phys.}{22}{2020}{013052}.

\bibitem{Holubec2020}
\Name{Holubec V. \etal} \REVIEW{Phys. Rev. Research}{2}{2020}{043262}.

\bibitem{Zakine2017}
\Name{Zakine R. \etal} \REVIEW{Entropy}{19}{2017}{193}.

\bibitem{Park2020}
\Name{Lee J.~S. \etal} \REVIEW{Phys. Rev. E}{102}{2020}{032116}.

\bibitem{Roldan2020}
\Name{Gopal A. \etal} \REVIEW{J. Phys. A: Math. Theor.}{54}{2021}{164001}.

\bibitem{Saha2020}
\Name{Kumari A. \etal} \REVIEW{Phys. Rev. E}{101}{2020}{032109}.

\bibitem{Speck2018}
\Name{Speck T.} \REVIEW{{EPL}}{123}{2018}{20007}.

\bibitem{Ekeh2020}
\Name{Ekeh T. \etal} \REVIEW{Phys. Rev. E}{102}{2020}{010101}.

\bibitem{Szamel2020}
\Name{Szamel G.} \REVIEW{Phys. Rev. E}{102}{2020}{042605}.

\bibitem{Stark2020}
\Name{Malgaretti P. \etal} \REVIEW{arXiv e-print}{}{2020}{2008.05257}.

\bibitem{Saha2019}
\Name{Saha A. \etal} \REVIEW{J. Stat. Mech.}{2019}{2019}{094012}.

\bibitem{Holubec2020a}
\Name{Holubec V. \etal} \REVIEW{Phys. Rev. E}{102}{2020}{060101}.

\bibitem{Seifert2008}
\Name{Schmiedl T. \etal} \REVIEW{EPL}{81}{2008}{20003}.

\bibitem{Esposito2009}
\Name{Esposito M. \etal} \REVIEW{Phys. Rev. Lett.}{102}{2009}{130602}.

\bibitem{Gingrich2016}
\Name{Gingrich T.~R. \etal} \REVIEW{Phys. Rev. Lett.}{116}{2016}{120601}.

\bibitem{Koyuk2019}
\Name{Koyuk T. \etal} \REVIEW{Phys. Rev. Lett.}{122}{2019}{230601}.

\bibitem{Mahmud2009}
\Name{Mahmud G. \etal} \REVIEW{Nat. Phys.}{5}{2009}{606}.

\bibitem{Maggi2014}
\Name{Koumakis N. \etal} \REVIEW{Soft Matter}{10}{2014}{5695}.

\bibitem{Yariv2014}
\Name{Yariv E. \etal} \REVIEW{Phys. Rev. E}{90}{2014}{032115}.

\bibitem{Martin2020}
\Name{Martin D. \etal} \REVIEW{Phys. Rev. E}{103}{2021}{032607}.

\bibitem{Reichhardt2008}
\Name{Wan M.~B. \etal} \REVIEW{Phys. Rev. Lett.}{101}{2008}{018102}.

\bibitem{Angelani2009}
\Name{Angelani L. \etal} \REVIEW{Phys. Rev. Lett.}{102}{2009}{048104}.

\bibitem{Reichhardt2013}
\Name{Reichhardt C. \etal} \REVIEW{Phys. Rev. E}{88}{2013}{062310}.

\bibitem{Reichhardt2018}
\Name{Reichhardt C. \etal} \REVIEW{Phys. Rev. E}{97}{2018}{052613}.

\bibitem{Sood2016}
\Name{Krishnamurthy S. \etal} \REVIEW{Nat. Phys.}{12}{2016}{1134}.

\bibitem{Blickle2011}
\Name{Blickle V. \etal} \REVIEW{Nat. Phys.}{8}{2011}{143}.

\bibitem{Martinez2016}
\Name{Mart\'inez I.~A. \etal} \REVIEW{Nat. Phys.}{12}{2016}{67}.

\bibitem{Leonardo2017}
\Name{Vizsnyiczai G. \etal} \REVIEW{Nat. Commun.}{8}{2017}{15974}.

\bibitem{Arlt2018}
\Name{Arlt J. \etal} \REVIEW{Nat. Commun.}{9}{2018}{768}.

\bibitem{Martin2018}
\Name{Martin D. \etal} \REVIEW{{EPL}}{121}{2018}{60005}.

\bibitem{Wulfert2017}
\Name{Wulfert R. \etal} \REVIEW{Phys. Rev. E}{95}{2017}{050103}.

\bibitem{Reichhardt2012}
\Name{Drocco J.~A. \etal} \REVIEW{Phys. Rev. E}{85}{2012}{056102}.

\bibitem{Valeriani2020}
\Name{Martinez R. \etal} \REVIEW{Soft Matter}{16}{2020}{4739}.

\bibitem{Yeomans2016}
\Name{Thampi S.~P. \etal} \REVIEW{Sci. Adv.}{2}{2016}{e1501854}.

\bibitem{Markovich2020}
\Name{Markovich T. \etal} \REVIEW{arXiv e-print}{}{2020}{2008.06735}.

\end{thebibliography}

\end{document}